\documentclass[12pt]{article}

\catcode`\@=11
\@addtoreset{equation}{section}

\global\arraycolsep=1pt
\def\fnote#1#2{\begingroup\def\thefootnote{#1}\footnote{#2}\addtocounter
{footnote}{-1}\endgroup}

\usepackage{amsbsy,amssymb,latexsym,amsfonts}

\newcommand{\beq}{\begin{equation}}
\newcommand{\eeq}{\end{equation}}
\newcommand{\beqa}{\begin{eqnarray}}
\newcommand{\eeqa}{\end{eqnarray}}

\newcommand{\CL}{{\mathcal L}}
\newcommand{\CD}{{\mathcal D}}
\newcommand{\CF}{{\mathcal F}}

\newcommand\fD{\mathfrak D}

\newcommand\langled{\langle\!\langle}
\newcommand\rangled{\rangle\!\rangle}

\def\llangle{\langle \hspace{-1mm} \langle}
\def\rrangle{\rangle \hspace{-1mm} \rangle}

\begin{document}
\begin{flushright}
OCU-PHYS 254\\
hep-th/0609039 \\
\end{flushright}
\bigskip

\begin{center}
{\bf\Large
Partial Breaking of $\mathcal{N}=2$ Supersymmetry \\
and Decoupling Limit of Nambu-Goldstone Fermion \\
in $U(N)$ Gauge Model\\
}

\bigskip\bigskip
{
K. Fujiwara\fnote{$*$}{e-mail: \texttt{fujiwara@sci.osaka-cu.ac.jp}}
}
\end{center}

\bigskip

\begin{center}
{\it Department of Mathematics and Physics,
Graduate School of Science\\
Osaka City University\\
3-3-138, Sugimoto, Sumiyoshi-ku, Osaka, 558-8585, Japan }

\end{center}

\vfill

\begin{abstract}
We study the $\mathcal{N}=1$ $U(N)$ gauge model obtained by
spontaneous breaking of $\mathcal{N}=2$ supersymmetry. 
The Fayet-Iliopoulos term included in the $\mathcal{N}=2$ action does not
appear in the resulting $\mathcal{N}=1$ action
and the superpotential
is modified to break discrete $R$ symmetry.
We take a limit in which the K\"ahler metric becomes flat
and the superpotential preserves non-trivial form.
The Nambu-Goldstone fermion is decoupled from other fields
but
the resulting action is still $\mathcal{N}=1$ supersymmetric.
It shows the origin of the fermionic shift symmetry
in $\mathcal{N}=1$ $U(N)$ gauge theory.
\end{abstract}

\newpage
\section{Introduction}

It was conjectured in \cite{DV} that non-perturbative quantities in
a low energy effective gauge theory can be computed by a matrix model.
This conjecture was confirmed by \cite{CDSW} for the case of an $\mathcal{N}=1$ $U(N)$ gauge theory
with a chiral superfield $\Phi$ in the adjoint representation of $U(N)$ .
The $\mathcal{N}=1$ action is obtained from ``soft" breaking of $\mathcal{N}=2$ supersymmetry
by adding the tree-level superpotential
\begin{eqnarray}
\int d^2 \theta \textrm{Tr} W(\Phi).
\end{eqnarray}
The group $SU(N)$ is confined and 
there is a symmetry of shifting the $U(1)$ gaugino by an anticommuting c-number
$\mathcal{W}_{\alpha} \rightarrow \mathcal{W}_{\alpha}-4\pi \chi_{\alpha}\ $.
It is called ``fermionic shift symmetry".
Thanks to this symmetry, effective superpotential is written as
\begin{eqnarray}
W_{\textrm{eff}}=\int d^2 \chi \mathcal{F},
\end{eqnarray} 
for some function $\mathcal{F}$. 
The fermionic shift symmetry 
is due to a free fermion and
should be related to a second, spontaneously broken supersymmetry.

Antoniadis-Partouche-Taylor (APT) constructed the $U(1)$ gauge model which breaks $\mathcal{N}=2$ supersymmetry to $\mathcal{N}=1$ spontaneously
by electric and magnetic Fayet-Iliopoulos (FI) terms \cite{APT}.
(See also \cite{PP}.)
The $U(N)$ generalization of the APT model was given in \cite{FIS1}
which is
described by $\mathcal{N}=1$ chiral superfields and $\mathcal{N}=1$ vector superfields.
The Nambu-Goldstone fermion appears in the overall $U(1)$ part of $U(N)$ gauge group
and couples with the $SU(N)$ sector
because of the fact that the 3rd derivatives of the prepotential are non-vanishing \cite{FIS3}
.
A manifestly $\mathcal{N}=2$ formulation 
of $U(N)$ gauge model \cite{FIS1,FIS3} with/without $\mathcal{N}=2$ hypermultiplets 
has been realized in \cite{FIS4}.
It overcomes the difficulty in coupling hypermultiplets to the APT model.
Partial breaking of local $\mathcal{N}=2$ supersymmetry was discussed in a lot of papers \cite{local,IM}.

This paper is organized as follows.
In section \ref{resultingN=1action}, 
We derive the resulting $\mathcal{N}=1$ $U(N)$ action from the $\mathcal{N}=2$ $U(N)$ gauge model \cite{FIS1}.
In section \ref{scaling limit}, 
we take a limit in which the K\"ahler metric becomes flat ,
while the superpotential preserves its non-trivial form.
After taking this limit the Nambu-Goldstone fermion is decoupled from other fields, but partial breaking of $\mathcal{N}=2$
supersymmetry is realized as before.
We get a general $\mathcal{N}=1$ action discussed in \cite{DV,CDSW}.
It shows that the fermionic shift symmetry is due to the decoupling limit of the Nambu-Goldstone fermion. 
In the appendix, we derive the resulting $\mathcal{N}=1$ supercharge algebra
\footnote
{We follow the notation of \cite{WB}}
.

\section{Spontaneous partial breaking of $\mathcal{N}=2$ supersymmetry and resulting $\mathcal{N}=1$ action} \label{resultingN=1action}

The on-shell action of the $\mathcal{N}=2$ $U(N)$ gauge model \cite{FIS1}
takes the following form:
\begin{eqnarray}
\label{L}
\CL_{\mathcal{N}=2 \atop \rm{on-shell}}
&=&
\CL_{\rm{kin}}
+\CL_{\rm{pot}}
+\CL_{\rm{Pauli}}
+\CL_{\rm{mass}}
+\CL_{\rm{fermi^4}}
~, \label{onshellN=2action}
\end{eqnarray}
with
\begin{eqnarray}
\CL_{\rm{kin}}&=&
-g_{ab}\mathcal{D}_m A^a \mathcal{D}^m A^{*b}
-\frac{1}{4} g_{ab} v_{mn}^a v^{bmn}
-\frac{1}{8} \textrm{Re}(\mathcal{F}_{ab}) \epsilon^{mnpq}v_{mn}^a v_{pq}^b 
\\&&
-\frac{1}{2} \mathcal{F}_{ab} \lambda^a \sigma^m \mathcal{D}_m
\bar{\lambda}^b
-\frac{1}{2} {\mathcal{F}}^*_{ab} \mathcal{D}_m \lambda^a
\sigma^m \bar{\lambda}^b
-\frac{1}{2} \mathcal{F}_{ab} \psi^a \sigma^m \mathcal{D}_m
\bar{\psi}^b
-\frac{1}{2} {\mathcal{F}}^*_{ab} \mathcal{D}_m \psi^a
\sigma^m \bar{\psi}^b ,
\nonumber \\
\CL_{\rm{pot}}&=&
-\frac{1}{2} g^{ab} \left(
\frac{1}{2}\fD_{a} + \sqrt{2} \xi\delta_{a}^0
\right)
\left(
\frac{1}{2}\fD_{b} + \sqrt{2} \xi\delta_{b}^0
\right)
-g^{ab}\partial_a W \partial_{b^*} {W}^*,
\\
\CL_{\rm{Pauli}}&=&
i\frac{\sqrt{2}}{8} \mathcal{F}_{abc} \psi^c \sigma^m
\bar{\sigma}^n \lambda^a v_{mn}^b
+i\frac{\sqrt{2}}{8} \mathcal{F}^*_{abc} \bar{\lambda}^a \bar{\sigma}^m \sigma^n
\bar{\psi}^cv_{mn}^b,
\\
\CL_{\rm{mass}}&=&
\left( -\frac{i}{4} \mathcal{F}_{abc} g^{cd} \partial_d W -\frac{1}{2} \partial_a \partial_b W \right) \psi^a \psi^b
-\frac{i}{4} \mathcal{F}_{abc} g^{cd} \partial_{d^*} W^* \lambda^a \lambda^b
\nonumber \\&&
+
\left\{ 
-\frac{1}{4\sqrt{2}} \mathcal{F}_{abc} g^{cd} \left( \mathfrak{D}_d +2\sqrt{2} \xi \delta^o_d \right)
+\frac{1}{\sqrt{2}} g_{ac}k_b^{*c}
\right\}
\psi^a \lambda^b
+c.c. \ ,
\\
\CL_{\rm{fermi^4}}&=&
-\frac{i}{8}\mathcal{F}_{abcd}
\psi^c \psi^d \lambda^a\lambda^b
+\frac{i}{8} \mathcal{F}^*_{abcd} \bar{\psi}^c\bar{\psi}^d\bar{\lambda}^a\bar{\lambda}^b
+g_{ab} \hat{F}^a \hat{F}^{*b} +\frac{1}{2} g_{ab} \hat{D}^a \hat{D}^b
\nonumber \\&&
+\frac{i}{4} \mathcal{F}_{abc} \hat{F}^{*c} \psi^a \psi^b 
+\frac{i}{4}\mathcal{F}_{abc} \hat{F}^c \lambda^a \lambda^b 
+\frac{1}{2\sqrt{2}} \mathcal{F}_{abc}\hat{D}^c \psi^a \lambda^b
\nonumber \\&&
-\frac{i}{4} \mathcal{F}^*_{abc} \hat{F}^{c} \bar{\psi}^a \bar{\psi}^b 
-\frac{i}{4}\mathcal{F}^*_{abc} \hat{F}^{*c} \bar{\lambda}^a \bar{\lambda}^b 
+\frac{1}{2\sqrt{2}} \mathcal{F}^*_{abc}\hat{D}^c \bar{\psi}^a \bar{\lambda}^b .
\end{eqnarray}
where
$
\hat{D}^a\equiv
-\frac{\sqrt{2}}{4}g^{ab} 
\left( 
\CF_{bcd}
\psi^d\lambda^c
+
\CF_{bcd}^*
\bar\psi^d\bar\lambda^c \right)
,\ 
\hat{F}^a\equiv
\frac{i}{4}g^{ab} 
\left( 
\CF_{bcd}^*
\bar\lambda^c\bar\lambda^d
-
\CF_{bcd}
\psi^c\psi^d \right)
$
and 
$W=eA^0+m\mathcal{F}_0$.
Let us examine the case with 
$
\displaystyle{
\mathcal{F}=\sum_{k=0}^{n} \textrm{tr} \frac{g_k}{k!} \Phi^k .
}
\ 
$
The vacuum condition 
$
\displaystyle{
\frac{\partial \mathcal{L}_{\textrm{pot}}}{\partial A^a}=0
}
$ reduces to 

\vspace{-0.5cm}
\begin{eqnarray}
\langle 
\mathcal{F}_{00} 
\rangle
= \frac{-e\pm i \xi}{m}, \label{VEV}
\end{eqnarray}
where $\langle$...$\rangle$ denotes ... evaluated at $A^r=0$ (indices $r$ represent non-Cartan generators).
For the sake of simplicity 
, we choose + sign in (\ref{VEV})
and this means $\frac{\xi}{m}\geq 0$.
It is revealed in \cite{FIS3} that the Nambu-Goldstone fermion exists in the overall $U(1)$ part of $U(N)$ gauge group,
\begin{eqnarray}
\llangle \delta_{\mathcal{N}=2} \left( \frac{\lambda^0-\psi^0}{\sqrt{2}} \right) \rrangle
= -2im (\eta_1 + \eta_2) ,
\ \ 
\llangle \delta_{\mathcal{N}=2} \left( \frac{\lambda^0+\psi^0}{\sqrt{2}} \right) \rrangle
= 0 .
\end{eqnarray}
We use $\llangle$...$\rrangle$ for vacuum expectation values which satisfy (\ref{VEV}).
$\frac{\lambda^0-\psi^0}{\sqrt{2}}$ is the Nambu-Goldstone fermion and it will be included in 
the overall $U(1)$ part of the resulting $\mathcal{N}=1$ $U(N)$ vector superfield.
The vacuum expectation value of the scalar potential $\mathcal{V}\equiv -\mathcal{L}_{\textrm{pot}}$ is
$\llangle \mathcal{V} \rrangle = 2m\xi$.
As is pointed out in \cite{FIS1}, the second term in the RHS of the local version of 
$\mathcal{N}=2$ supersymmetry algebra
enables us to
add a constant $2m\xi$ to the action (\ref{onshellN=2action})
in order to set 
$\langled \mathcal{V} \rangled=0$.
In the formalism of harmonic superspace,
this freedom to add a constant number comes from 
arbitrariness to choose the imaginary part of the magnetic 
FI term in \cite{FIS4}
\footnote
{
In \cite{APT}, such freedom comes from the electric FI term.
}.

To obtain the resulting $\mathcal{N}=1$ action 
for the case that U(N) gauge symmetry is not broken at vacua,
we shift the scalar fields $A^a$ by its vacuum expectation value
and mix
the spinor fields $\psi^a$ and $\lambda^a$.
We define
\begin{eqnarray}
\tilde{A}^a \equiv A^a - \langled A^0  \rangled \delta_0^a ,\ 
\lambda^{-a}
\equiv
\frac{1}{\sqrt{2}} (\lambda^{a}-\psi^{a}),\ 
\lambda^{+a}
\equiv
\frac{1}{\sqrt{2}} (\lambda^{a}+\psi^{a}) \ \ .
\end{eqnarray}
Substitute these into (\ref{onshellN=2action}), we get
the resulting $\mathcal{N}=1$ $U(N)$ gauge action after spontaneous breaking of $\mathcal{N}=2$ supersymmetry,
\begin{eqnarray}
\CL_{\mathcal{N}=1 \atop \rm{on-shell}}
=\tilde{\CL}_{\rm{kin}} + \tilde{\CL}_{\rm{pot}} + \tilde{\CL}_{\rm{Pauli}} + \tilde{\CL}_{\rm{mass}} + \tilde{\CL}_{\rm{fermi}^4}
 ,
\label{onshellN=1action}
\end{eqnarray}
with
\begin{eqnarray}
&&\tilde{\CL}_{\rm{kin}}=
-\tilde{g}_{ab}\mathcal{D}_m \tilde{A}^a \mathcal{D}^m \tilde{A}^{*b}
-\frac{1}{4} \tilde{g}_{ab} v_{mn}^a v^{bmn}
-\frac{1}{8} \textrm{Re}(\tilde{\mathcal{F}}_{ab}) \epsilon^{mnpq}v_{mn}^a v_{pq}^b
\\&& \ \ \ \ 
-\frac{1}{2} \tilde{\mathcal{F}}_{a b}
\lambda^{-a} \sigma^m \mathcal{D}_m \bar{\lambda}^{-b}
-\frac{1}{2} \tilde{\mathcal{F}}^*_{a b}
\mathcal{D}_m \lambda^{-a} \sigma^m \bar{\lambda}^{-b}
-\frac{1}{2} \tilde{\mathcal{F}}_{a b}
\lambda^{+a} \sigma^m \mathcal{D}_m \bar{\lambda}^{+b}
-\frac{1}{2} \tilde{\mathcal{F}}^*_{a b}
\mathcal{D}_m \lambda^{+a} \sigma^m \bar{\lambda}^{+b} ,
\nonumber \\&&
\tilde{\CL}_{\rm{pot}}=
-\frac{1}{8} \tilde{g}^{ab} \tilde{\mathfrak{D}}_a \tilde{\mathfrak{D}}_b
-\tilde{g}^{ab} \tilde{\partial}_a \widetilde{W} \tilde{\partial}_{b^*} \widetilde{W}^* ,
\\&&
\tilde{\CL}_{\rm{Pauli}}=
i\frac{\sqrt{2}}{8} \tilde{\mathcal{F}}_{abc} \lambda^{+c} \sigma^m
\bar{\sigma}^n \lambda^{-a} v_{mn}^b
+i\frac{\sqrt{2}}{8} \tilde{\mathcal{F}}^*_{abc} \bar{\lambda}^{-a} \bar{\sigma}^m \sigma^n
\bar{\lambda}^{+c}v_{mn}^b
\\&&
\tilde{\CL}_{\rm{mass}}=
\left( -\frac{i}{4} \tilde{\mathcal{F}}_{abc} \tilde{g}^{cd} \tilde{\partial}_d \widetilde{W} -\frac{1}{2} \tilde{\partial}_a \tilde{\partial}_b \widetilde{W} \right) \lambda^{+a} \lambda^{+b}
-\frac{i}{4} \tilde{\mathcal{F}}_{abc} \tilde{g}^{cd} \tilde{\partial}_{d^*} \widetilde{W}^* \lambda^{-a} \lambda^{-b}
\nonumber \\&& \ \ \ \ \ \ \ \ \ \ 
+
\left\{ 
-\frac{1}{4\sqrt{2}} \tilde{\mathcal{F}}_{abc} \tilde{g}^{cd} \tilde{\mathfrak{D}}_d
+\frac{1}{\sqrt{2}} \tilde{g}_{ac}\tilde{k}_b^{*c}
\right\}
\lambda^{+a} \lambda^{-b}
+c.c. \ ,
\\&&
\tilde{\CL}_{\rm{fermi^4}}=
-\frac{i}{8}\tilde{\mathcal{F}}_{abcd}
\lambda^{+c} \lambda^{+d} \lambda^{-a}\lambda^{-b}
+\frac{i}{8} \tilde{\mathcal{F}}^*_{abcd} \bar{\lambda}^{+c}\bar{\lambda}^{+d}\bar{\lambda}^{-a}\bar{\lambda}^{-b}
+\tilde{g}_{ab} \check{F}^a \check{F}^{*b} +\frac{1}{2} \tilde{g}_{ab} \check{D}^a \check{D}^b
\nonumber \\&&  \ \ \ \ \ \ \ \ \ \ \ 
+\frac{i}{4} \tilde{\mathcal{F}}_{abc} \check{F}^{*c} \lambda^{+a} \lambda^{+b} 
+\frac{i}{4}\tilde{\mathcal{F}}_{abc} \check{F}^c \lambda^{-a} \lambda^{-b} 
+\frac{1}{2\sqrt{2}} \tilde{\mathcal{F}}_{abc}\check{D}^c \lambda^{+a} \lambda^{-b}
\nonumber \\&&  \ \ \ \ \ \ \ \ \ \ \ 
-\frac{i}{4} \tilde{\mathcal{F}}^*_{abc} \check{F}^{c} \bar{\lambda}^{+a} \bar{\lambda}^{+b} 
-\frac{i}{4}\tilde{\mathcal{F}}^*_{abc} \check{F}^{*c} \bar{\lambda}^{-a} \bar{\lambda}^{-b} 
+\frac{1}{2\sqrt{2}} \tilde{\mathcal{F}}^*_{abc}\check{D}^c \bar{\lambda}^{+a} \bar{\lambda}^{+b} .
\end{eqnarray}
where
\begin{eqnarray}
&&
\tilde{\mathcal{F}}(\tilde{A})
\equiv
\langled \mathcal{F} \rangled 
+\langled \mathcal{F}_a \rangled \tilde{A}^a 
+\langled \mathcal{F}_{ab} \rangled \frac{\tilde{A}^a \tilde{A}^b}{2!} 
+\langled \mathcal{F}_{abc} \rangled \frac{\tilde{A}^a \tilde{A}^b \tilde{A}^c}{3!}
+\cdots , \ 
\tilde{\mathcal{F}}_a
\equiv 
\frac{\partial \tilde{\mathcal{F}}(\tilde{A})}{\partial \tilde{A}^a}
,
\nonumber
\\
&&
 \tilde{\mathcal{F}}_{ab}
\equiv  
\frac{\partial^2 \tilde{\mathcal{F}}}{\partial \tilde{A}^a \partial \tilde{A}^b}, \ \cdots ,\ 
\tilde{g}_{ab}
\equiv
\frac{\tilde{\mathcal{F}}_{ab}-\tilde{\mathcal{F}}_{ab}^*}{2i}  
,\ 
\tilde{\mathfrak{D}}_a 
\equiv
-i\tilde{g}_{ab} f^b_{cd} \tilde{A}^{*c} \tilde{A}^d
,\ 
\tilde{k}_a^{\ b} \equiv 
-i\tilde{g}^{bc} \frac{\partial}{\partial \tilde{A}^{*c}} \tilde{\mathfrak{D}}_a,
\nonumber
\\
&&
\check{F}^a
\equiv
\frac{i}{4}\tilde{g}^{ab} \tilde{\mathcal{F}}_{bcd}^* \bar{\lambda}^{-c} \bar{\lambda}^{-d} 
-\frac{i}{4}\tilde{g}^{ab} \tilde{\mathcal{F}}_{bcd} \lambda^{+c} \lambda^{+d}
,\ 
\check{D}^a
\equiv
-\frac{\sqrt{2}}{4} \tilde{g}^{ab} \tilde{\mathcal{F}}_{bcd} \lambda^{+c} \lambda^{-d}
-\frac{\sqrt{2}}{4} \tilde{g}^{ab} \tilde{\mathcal{F}}_{bcd}^* \bar{\lambda}^{+c} \bar{\lambda}^{-d},
\nonumber
\\
&&
\widetilde{W}
\equiv
(e-i\xi)\tilde{A}^0+m\tilde{\mathcal{F}}_0 ,\ 
\tilde{\partial}_a \widetilde{W}
\equiv
\frac{\partial \widetilde{W}}{\partial \tilde{A}^a} ,\ 
\tilde{\partial}_a \tilde{\partial}_b \widetilde{W} 
\equiv
\frac{\partial^2 \widetilde{W}}{\partial \tilde{A}^a \partial \tilde{A}^b} \ 
.\nonumber
\end{eqnarray}

Here we have used
\begin{eqnarray}
i\partial_a \mathfrak{D}_b+i\partial_b \mathfrak{D}_a-\frac{1}{2} g^{cd} \mathcal{F}_{abc} \mathfrak{D}_d=0
,\ \ 
g^{ab} \mathfrak{D}_a \delta_b^0=0
, \\
\mathcal{F}_{abc} \lambda^{+a} \sigma^n \bar{\sigma} ^m \lambda^{+b} v_{mn}^c=0,
\ \ 
\mathcal{F}_{abcd}\lambda^{+a}\lambda^{+b}\lambda^{+c}\lambda^{+d}=0.
\end{eqnarray}
Take notice that we have added the constant $2m\xi$ to $\mathcal{L}_{\textrm{pot}}$ as mentioned above.

As a result, the action (\ref{onshellN=1action}) agrees with the action (\ref{onshellN=2action}) except for the superpotential term and FI term.
There is no FI term in (\ref{onshellN=1action}), and the superpotential $W=eA^o +m\mathcal{F}_0$ get shifted to $\widetilde{W}=(e-i\xi)\tilde{A}^0+m\tilde{\mathcal{F}}_0$ (we neglected a constant term).
Because the coefficient $(e-i\xi)$ in $\widetilde{W}$ is a complex number, (\ref{onshellN=1action}) is not invariant under the discrete $R$ transformation
\footnote
{\lefteqn{
$$
R:
\left(
\begin{array}{c}
\lambda^{-a} \\
\lambda^{+a}
\end{array}
\right)
\longrightarrow 
\left(
\begin{array}{c}
\lambda^{+a} \\
-\lambda^{-a}
\end{array}
\right)
$$
}}
.

We can write the off-shell $\mathcal{N}=1$ action
by introducing
auxiliary fields $\tilde{F}$ and $\tilde{D}$,
\begin{eqnarray}
\lefteqn{\CL_{\mathcal{N}=1 \atop \rm{off-shell}}=}
\nonumber \\
&&
-\tilde{g}_{ab}\CD_m\tilde{A}^a\CD^m\tilde{A}^{*b}
-\frac{1}{4}
\tilde{g}_{ab}v_{mn}^av^{bmn}
-\frac{1}{8}
{\rm Re}
(\tilde{\CF}_{ab})\epsilon^{mnpq}v_{mn}^av_{pq}^b
\nonumber\\&&
-\frac{1}{2}\tilde{\CF}_{ab}\lambda^{-a}\sigma^m\CD_m\bar\lambda^{-b}
-\frac{1}{2}\tilde{\CF}_{ab}^*\CD_m\lambda^{-a}\sigma^m\bar\lambda^{-b}
-\frac{1}{2}\tilde{\CF}_{ab}\lambda^{+a}\sigma^m\CD_m\bar\lambda^{+b}
-\frac{1}{2}\tilde{\CF}_{ab}^*\CD_m\lambda^{+a}\sigma^m\bar\lambda^{+b}
\nonumber\\&&
+\tilde{g}_{ab}\tilde{F}^a\tilde{F}^{*b}
+\tilde{F}^a\tilde{\partial}_a \widetilde{W}
+\tilde{F}^{*a}\tilde{\partial}_{a^*}  \widetilde{W}^*
+\frac{1}{2}
\tilde{g}_{ab}\tilde{D}^a\tilde{D}^b
+\frac{1}{2}\tilde{D}^a\tilde{\fD}_a
\nonumber\\&&
+(\frac{i}{4}\tilde{\CF}_{abc}\tilde{F}^{*c}-\frac{1}{2}\tilde{\partial}_a\tilde{\partial}_b \widetilde{W})
 \lambda^{+a}\lambda^{+b}
+\frac{i}{4}\tilde{\CF}_{abc}
\tilde{F}^c\lambda^{-a}\lambda^{-b} 
+\frac{1}{\sqrt{2}}(\tilde{g}_{ac}k_{b}^*{}^{c}+\frac{1}{2}\tilde{\CF}_{abc}\tilde{D}^c)
\lambda^{+a}\lambda^{-b}
\nonumber\\&&
+(-\frac{i}{4}\tilde{\CF}^*_{abc}\tilde{F}^{c}-\frac{1}{2}\tilde{\partial}_{a^*}
\tilde{\partial}_{b^*}  \widetilde{W}^*)
\bar\lambda^{+a}\bar\lambda^{+b} 
-\frac{i}{4}\tilde{\CF}^*_{abc}
\tilde{F}^{*c}\bar\lambda^{-a}\bar\lambda^{-b}
+\frac{1}{\sqrt{2}}(\tilde{g}_{ca}k_{b}{}^{c}+\frac{1}{2}\tilde{\CF}^*_{abc}\tilde{D}^c)
\bar\lambda^{+a}\bar\lambda^{-b}
\nonumber\\&&
-i\frac{\sqrt{2}}{8}(
\tilde{\CF}_{abc}
\lambda^{+c}\sigma^n\bar\sigma^m\lambda^{-a}
-\tilde{\CF}^*_{abc}
\bar\lambda^{-a}\bar\sigma^m\sigma^n\bar\lambda^{+c}
)v_{mn}^b
\nonumber\\&&
-\frac{i}{8}\tilde{\CF}_{abcd}
\lambda^{+c}\lambda^{+d}\lambda^{-a}\lambda^{-b}
+\frac{i}{8}\tilde{\CF}^*_{abcd}
\bar\lambda^{+c}\bar\lambda^{+d}\bar\lambda^{-a}\bar\lambda^{-b}. \label{offshellN=1action}
\end{eqnarray}
Component fields ($\tilde{A}^a$, $\lambda^{+a}$, $\tilde{F}^a$) form massive $\mathcal{N}=1$ chiral
multiplets $\tilde{\Phi}^a$.
Other component fields ($v_m^a$, $\lambda^{-a}$, $\tilde{D}^a$) form massless $\mathcal{N}=1$ vector
multiplets $\tilde{V}^a$.
The Nambu-Goldstone fermion $\lambda^{-0}$ is contained
in the overall $U(1)$ part of $\tilde{V}^a$.

\section{Reparametrization and scaling limit} \label{scaling limit}

We consider a limit in which the Nambu-Goldstone fermion $\lambda^{-0}$ is decoupled from other fields 
with $\mathcal{N}=2$ supersymmetry breaking to $\mathcal{N}=1$. 
If the prepotential $\mathcal{F}$ is a second order polynomial, 
there are no Yukawa couplings in (\ref{offshellN=1action}) and $\lambda^{-0}$ will be a free fermion.
However, derivatives of the superpotential become zero,
$\tilde{\partial}_a \tilde{\partial}_b \widetilde{W}=m\tilde{\mathcal{F}}_{0ab}=0$ and
$\tilde{\partial}_a \widetilde{W}=(e-i\xi)\delta_a^0 + m \tilde{\mathcal{F}}_{0a}=(e-i\xi)\delta_a^0+m \langled \mathcal{F}_{0a} \rangled=0$.
This means that the superpotential does not contribute to (\ref{offshellN=1action}) and it preserves $\mathcal{N}=2$ supersymmetry.
This problem can be solved by a large limit of the parameters $(e, m, \xi)$, \ 
i.e. large limit of electric and magnetic FI terms.

We reparametrize $g_k = \frac{g'_k}{\Lambda}(k \geq 3)$ and $(e,\ m,\ \xi) = (\Lambda e',\ \Lambda m',\ \Lambda \xi')$.
The prepotential $\mathcal{F}$ is 
\begin{eqnarray}
&& \mathcal{F}=\sum_{k=0}^{n} \textrm{tr} \frac{g_k}{k!} \Phi^k 
=
\textrm{tr}
\left(
g_0 \bold{1} +g_1 \Phi+\frac{g_2}{2} \Phi^2
\right)
+
\frac{1}{\Lambda} \sum_{k=3}^{n} \textrm{tr} \frac{g'_k}{k!} \Phi^k,
\end{eqnarray}
and we see the $\Lambda$ dependence of the following terms:
\begin{eqnarray}
&&
\tilde{\mathcal{F}}_{ab}
=
\langled
\mathcal{F}_{ab}
\rangled
+
\frac{1}{\Lambda}
\left\{
\langled
\mathcal{F}'_{abc}
\rangled 
\tilde{A}^c
+
\langled
\mathcal{F}'_{abcd}
\rangled 
\frac{
\tilde{A}^c \tilde{A}^{d}
}
{2!}
+ \ \cdots
\right\}
=
\frac{-e+i\xi}{m} \delta_{ab}+\mathcal{O}(\Lambda^{-1}),\ 
\nonumber \\&&
\tilde{\mathcal{F}}_{abc}=\mathcal{O}(\Lambda^{-1}),\ 
\tilde{\mathcal{F}}_{abcd}=\mathcal{O}(\Lambda^{-1}),\ 
\tilde{g}_{ab} = \frac{\xi}{m} \delta_{ab}
+\mathcal{O}(\Lambda^{-1}) ,
\nonumber \\&&
\tilde{\mathfrak{D}}_a
=
-i\tilde{g}_{ab} f^b_{cd} \tilde{A}^{*c} \tilde{A}^d
=
-\frac{i\xi}{m} \delta_{ab} f^b_{cd} \tilde{A}^{*c} \tilde{A}^d
+\mathcal{O} (\Lambda^{-1}) \ .
\end{eqnarray}
where
$
\displaystyle{
\mathcal{F}'=
\textrm{tr}
\left(
g_0 \bold{1} +g_1 \Phi+\frac{g_2}{2} \Phi^2
\right)
+
\sum_{k=3}^{n} \textrm{tr} \frac{g'_k}{k!} \Phi^k
}$.\ 
Note that the scaling parameter $\Lambda$ is cancelled out in the superpotential term:
\begin{eqnarray}
\tilde{\partial}_a \widetilde{W}&=&
(e-i\xi)\delta_a^0 +m\tilde{\mathcal{F}}_{0a} 
=
m'
\left\{
\langled \mathcal{F}'_{0ab} \rangled \tilde{A}^b
+\frac{1}{2!} \langled \mathcal{F}'_{0abc} \rangled \tilde{A}^b \tilde{A}^c +\ \cdots
\right\} ,
\\
\tilde{\partial}_a \tilde{\partial}_b \widetilde{W}
&=&
m'
\left\{
\langled \mathcal{F}'_{0ab} \rangled
+
\langled \mathcal{F}'_{0abc} \tilde{A}^c \rangled
+
\frac{1}{2!} \langled \mathcal{F}'_{0abcd} \rangled \tilde{A}^c \tilde{A}^d+\ \cdots
\right\}.
\end{eqnarray}

Take a limit $\Lambda \rightarrow \infty$, and the action (\ref{offshellN=1action}) is converted into
\begin{eqnarray}
\CL&=&
\frac{\xi}{m}
\delta_{ab} \Biggl\{
-\CD_m\tilde{A}^a\CD^m\tilde{A}^{*b}
-i\lambda^{+a}\sigma^m\CD_m\bar\lambda^{+b}
\nonumber\\&&
+\tilde{F}^a\tilde{F}^{*b}
-\frac{i}{2} f^b_{cd} \tilde{D}^a \tilde{A}^{*c} \tilde{A}^d
+\frac{\sqrt{2}}{2} f^b_{dc} \tilde{A}^{*c}
\lambda^{+a}\lambda^{-d}
+\frac{\sqrt{2}}{2} f^b_{dc} \tilde{A}^{c}
\bar\lambda^{+a}\bar\lambda^{-d}
\Biggr\}
\nonumber\\&&
+
\frac{\xi}{m}
\delta_{ab} \Biggl\{
-\frac{1}{4}
v_{mn}^av^{bmn}
+\frac{1}{8}
\frac{e}{\xi} \epsilon^{mnpq}v_{mn}^av_{pq}^b
-i\lambda^{-a}\sigma^m\CD_m\bar\lambda^{-b}
+\frac{1}{2}
\tilde{D}^a \tilde{D}^b
\Biggr\}
\nonumber \\&&
+
\tilde{F}^a\tilde{\partial}_a \widetilde{W}
+\tilde{F}^{*a}\tilde{\partial}_{a^*}  \widetilde{W}^*
-\frac{1}{2}\tilde{\partial}_a\tilde{\partial}_b \widetilde{W} \lambda^{+a}\lambda^{+b}
-\frac{1}{2}\tilde{\partial}_{a^*} \tilde{\partial}_{b^*}  \widetilde{W}^* \bar\lambda^{+a}\bar\lambda^{+b} .
\label{classicalaction}
\end{eqnarray}
The matrix form of the superpotential $\widetilde{W}$ is given as
\footnote
{
We normalize the standard $u(N)$ Cartan generators $t_i$ as $\textrm{tr} (t_i t_j)=\frac{1}{2} \delta_{ij}$,
which implies that the overall $u(1)$ generator is $t_0=\frac{1}{\sqrt{2N}} \boldsymbol{1}_{N\times N}$ .
}
\begin{eqnarray}
\widetilde{W} &\equiv&
m'
\left\{
\frac{1}{2!}
\langled \mathcal{F}'_{0ab} \rangled \tilde{A}^a \tilde{A}^b
+\frac{1}{3!} \langled \mathcal{F}'_{0abc} \rangled \tilde{A}^a \tilde{A}^b \tilde{A}^c +\ \cdots
\right\}
\nonumber \\
&=&
m
\left\{
\frac{1}{2!}
\langled \mathcal{F}_{0ab} \rangled \tilde{A}^a \tilde{A}^b
+\frac{1}{3!} \langled \mathcal{F}_{0abc} \rangled \tilde{A}^a \tilde{A}^b \tilde{A}^c +\ \cdots
\right\}
\nonumber \\
&=&
\frac{m}{\sqrt{2N}}
\sum_{k=1}^n \frac{g_k}{(k-1)!} \textrm{tr} \left( \tilde{\boldsymbol{A}}+\frac{\langled A^0 \rangled}{\sqrt{2N}} \boldsymbol{1} \right)^{k-1}
-m\langled \mathcal{F}_0 \rangled
-m\langled \mathcal{F}_{0a} \rangled \tilde{A}^a
\nonumber \\
&=&
m
\sum_{k=1}^{n-2} \frac{h_k}{k+1} \textrm{tr} \tilde{\boldsymbol{A}}^{k+1} ,
\end{eqnarray}
where we define
$\displaystyle{h_k \equiv \frac{(k+1)}{\sqrt{2N}} \sum_{\ell=0}^{n-2-k} \frac{g_{k+\ell+2}}{(k+\ell+1)!} \ _{(k+\ell+1)}C_\ell \left( \frac{\langled A^0 \rangled}{\sqrt{2N}} \right)^\ell}$
.
Here the symbol $\ _{(k+\ell+1)} C_{\ell}$ is a binomial coefficient 
. 

We can rewrite the action (\ref{classicalaction}) in superfield formalism as
\begin{eqnarray}
\mathcal{L}
=
\textrm{Im}
\left[
\frac{-e+i\xi}{m}
\left(
2\int d^4 \theta \textrm{tr} \tilde{\Phi}^+ e^{\tilde{V}} \tilde{\Phi}
+
\int d^2 \theta \textrm{tr} \tilde{\mathcal{W}}^{\alpha} \tilde{\mathcal{W}}_{\alpha} 
\right)
\right]
+
\left(
\int d^2 \theta \widetilde{W}(\tilde{\Phi}) 
+c.c.
\right) ,\ \ \ \ \ 
\label{classicalaction2}
\end{eqnarray}
where $\tilde{\mathcal{W}}$ is the field strength of $\tilde{V}$.
The factor 2 in the first term comes from the normalization of the standard $u(N)$ Cartan generators. 
Note that the Nambu-Goldstone fermion $\lambda^{-0}$ , which is contained in the overall $U(1)$ part of 
$\mathcal{N}=1$ $U(N)$ vector superfields $\tilde{V}$,
is decoupled from other fields in (\ref{classicalaction2}) .
However $\mathcal{N}=2$ supersymmetry is broken to $\mathcal{N}=1$ because of existence of the superpotential.
We get a general $\mathcal{N}=1$ action (\ref{classicalaction2}), which is known as a ``soft" broken $\mathcal{N}=1$ action,
from spontaneously broken $\mathcal{N}=2$ supersymmetry.
We conclude that the fermionic shift symmetry in \cite{CDSW} is related to the decoupling limit of the Nambu-Goldstone fermion.

\section*{Acknowledgments}
The author would like to thank
Hiroshi Itoyama,
Makoto Sakaguchi,
Kazunobu Maruyoshi
and 
Hironobu Kihara
for very useful discussions.

\appendix
\section{Supercharge algebra}

The $\mathcal{N}=2$ transformation rule are given by a combination of following transformation rules
\footnote
{
It is easy to give proof that $\delta_{\eta_2} \mathcal{L}=0$ (up to total derivative) with the use of 
$\delta_{\eta_1} \mathcal{L}=0$ and $R \mathcal{L} = \mathcal{L} |_{\xi \rightarrow -\xi}$ \  .
(See \cite{FIS1}.)
As in \cite{Fayet}, the FI term does not break $\mathcal{N}=2$ supersymmetry.
}
,
\begin{eqnarray}
\left\{
\begin{array}{l}
\delta_{\eta_1} A^a=\sqrt{2} \eta_1 \psi^a
\nonumber \\
\delta_{\eta_1} \psi^a=i\sqrt{2}\sigma^m\bar\eta_1\CD_mA^a +\sqrt{2}\eta_1 (\hat F^a-g^{ab}\partial_{b^*} W^*)
\nonumber \\
\delta_{\eta_1} \lambda^a=\frac{1}{2}\sigma^m\bar\sigma^n\eta_1 v_{mn}^a +i\eta_1 (\hat D^a-\frac{1}{2}g^{ab}(\fD_b+2\sqrt{2} \xi \delta_b^{\ 0}))
\nonumber \\
\delta_{\eta_1} v_{m}^a =i\eta_1\sigma^m\bar\lambda^a -i\lambda^a\sigma^m\bar\eta_1
\end{array}
\right.
\\
\left\{
\begin{array}{l}
\delta_{\eta_2} A^a=- \sqrt{2}\eta_2 \lambda^a
\nonumber \\
\delta_{\eta_2} \psi^a=\frac{1}{2}\sigma^m\bar\sigma^n \eta_2 v_{mn}^a -i\eta_2(\hat D^a+\frac{1}{2}g^{ab}(\fD_b-2\sqrt{2} \xi \delta_b^{\ 0}))
\nonumber \\
\delta_{\eta_2} \lambda^a=-i\sqrt{2}\sigma^m\bar\eta_2\CD_mA^a -\sqrt{2}\eta_2 (\hat F^{*a}-g^{ab}\partial_{b^*} W^*)
\nonumber \\
\delta_{\eta_2} v_{m}^a =i\eta_2\sigma^m\bar\psi^a -i\psi^a\sigma^m\bar\eta_2
\end{array}
\right.
\end{eqnarray}
where spinors $\eta_k (k=1,2)$ are transformation parameters.
The $\mathcal{N}=2$ supersymmetric transformation rules are $\delta_{\mathcal{N}=2} \chi^a=\delta_{\eta_1} \chi^a + \delta_{\eta_2} \chi^a$.
We can find 
the 1st supercurrent $S_{1\alpha}^m$ from the action (\ref{onshellN=2action}):
\begin{eqnarray}
S^m_1&=&
-ig_{ab}  \sigma^{np} \sigma^m \bar{\lambda}^b v_{pn}^a
-\frac{1}{2}  \sigma^m \bar{\lambda}^a \mathfrak{D}_a
+i\sqrt{2} \left( 
e\delta_{c^*}^0+m\mathcal{F}_{0c}^* \right)  \sigma^m \bar{\psi}^c
\nonumber \\&&
-\sqrt{2} \xi  \sigma^m \bar{\lambda}^0
-\sqrt{2} g_{ab}  \sigma^n \bar{\sigma}^m \psi^a \mathcal{D}_n A^{*b} 
+\cdots ,
\end{eqnarray}
where the dots denote terms involving three fermions .
The 2nd supercurrent $S_{2\alpha}^m$ is given by the discrete $R$ transformation of $S_{1\alpha}^m$
with a flip of the sign of the FI parameter $\xi$,
\begin{eqnarray}
 S^m_2&=&
-ig_{ab}  \sigma^{np} \sigma^m \bar{\psi}^b v_{pn}^a
-\frac{1}{2}  \sigma^m \bar{\psi}^a \mathfrak{D}_a
-i\sqrt{2} \left( 
e\delta_{c^*}^0+m\mathcal{F}_{0c}^* \right)  \sigma^m \bar{\lambda}^c
\nonumber \\&&
+\sqrt{2} \xi  \sigma^m \bar{\psi}^0
+\sqrt{2} g_{ab}  \sigma^n \bar{\sigma}^m \lambda^a \mathcal{D}_n A^{*b} 
+\cdots .
\end{eqnarray}
Supercharge algebra is derived by
\begin{eqnarray}
\delta_{\eta_A} S_{B\alpha}^0
=
i
\left[
\eta_A Q_A+\bar{\eta}_A \bar{Q}_A,\ 
S^0_{B\alpha}
\right]
=
i \eta_A^{\beta}
\Bigl\{
Q_{A\beta},\ S_{B\alpha}^0
\Bigr\}
+
i \bar{\eta}_{A\dot{\beta}}
\Bigl\{
\bar{Q}_A^{\dot{\beta}},\ S_{B\alpha}^0
\Bigr\}
,
\end{eqnarray}
where $A,B$=1 or 2 .
It may be irrelevant to denote supercharges as $Q_1, Q_2$
because $\mathcal{N}=2$ supersymmetry is broken to $\mathcal{N}=1$ spontaneously and
the supercharge corresponding to the broken supersymmetry is ill-defined.
We ignore this point here
and write the divergent part explicitly.
We obtain the central charge
\begin{eqnarray}
\left\{
Q_{1\alpha},\ Q_{2\beta}
\right\}
&=&
\sqrt{2} i
\epsilon_{\beta \alpha}
\int d x^3
\partial_i
\left\{
\left(
A^{*b} \textrm{Re} \mathcal{F}_{ab} -2i \partial_a K
\right)
\epsilon^{0ijk} v_{jk}^a + 2g_{ab} A^{*b} v^{a0i}
\right\}
\nonumber 
\\&&
+8\xi 
\int d^3 x \partial_i 
\left\{
A^{*0}  (\sigma^{0i} \epsilon )_{\beta \alpha}  
\right\} .
\end{eqnarray}
Here $K=\frac{i}{2} (A^a \mathcal{F}_a^* - A^{*a} \mathcal{F}_a)$ is the K\"ahler potential.
To get the resulting $\mathcal{N}=1$ supercharge algebra, we define 
$Q^- \equiv \frac{1}{\sqrt{2}}(Q_1-Q_2)$
and 
$Q^+ \equiv \frac{1}{\sqrt{2}}(Q_1+Q_2)$
.
Anti-commutators of $Q^-$($Q^+$) and $\bar{Q}^-$($\bar{Q}^+$) are given as
\begin{eqnarray}
\left\{
Q^-_{\alpha},\ \bar{Q}^-_{\dot{\beta}}
\right\}
&=&
-i\int d^3 x 
\Bigl[
\frac{i}{4} g^{ab} (g_{ac}v^c_{np} \sigma^n \bar{\sigma}^p+i\mathfrak{D}_a)\sigma^0(g_{bd} v_{qr}^d \bar{\sigma}^q \sigma^r +i\mathfrak{D}_b)
\nonumber \\
&&
-2ig_{ab} \mathcal{D}_p A^a \mathcal{D}_n A^{*b} \sigma^n \bar{\sigma}^0 \sigma^p
-2ig^{ab} \partial_a \widetilde{W} \partial_{b^*} \widetilde{W}^* \sigma^0
+\cdots \ 
\Bigr]
_{\alpha \dot{\beta}} \ ,
\nonumber \\
\left\{
Q^+_{\alpha},\ \bar{Q}^+_{\dot{\beta}}
\right\}
&=&
-i\int d^3 x 
\Bigl[
\frac{i}{4} g^{ab} (g_{ac}v^c_{np} \sigma^n \bar{\sigma}^p+i\mathfrak{D}_a)\sigma^0(g_{bd} v_{qr}^d \bar{\sigma}^q \sigma^r +i\mathfrak{D}_b)
\nonumber \\
&&
-2ig_{ab} \mathcal{D}_p A^a \mathcal{D}_n A^{*b} \sigma^n \bar{\sigma}^0 \sigma^p
-2ig^{ab} \partial_a \widetilde{W} \partial_{b^*} \widetilde{W}^* \sigma^0
+\cdots \ 
\Bigr]
_{\alpha \dot{\beta}}
\nonumber \\
&&
-8m\xi \sigma^0_{\alpha \dot{\beta}}
\int d^3 x ,
\end{eqnarray}
where the dots indicate terms involving fermion fields.
This result agree with the supersymmetry algebra in \cite{Hughes}.
Finally, we conclude that $Q^-$ is the unbroken generator and $Q^+$ is the broken one.

\end{document}